\documentclass[aps,twocolumn,pra,superscriptaddress,floatfix,amsmath,amssymb,showpacs,tightenlines]{revtex4}
\usepackage{epsfig,graphicx,times}
\usepackage{amstext}
\usepackage{amsmath}
\usepackage{amssymb}
\usepackage{graphicx}
\usepackage{latexsym}
\usepackage{bm}
\begin{document}
\title{Scalable superconducting qubit circuits using
dressed states}

\author{ Yu-xi Liu}
\affiliation{Frontier Research System,  The Institute of
Physical and Chemical Research (RIKEN), Wako-shi, Saitama
351-0198, Japan}
\affiliation{CREST, Japan Science and
Technology Agency (JST), Kawaguchi, Saitama 332-0012, Japan}
\author{C. P. Sun}
\affiliation{Frontier Research System, The Institute of Physical
and Chemical Research (RIKEN), Wako-shi, Saitama 351-0198, Japan}
\affiliation{Institute of Theoretical Physics, The Chinese Academy
of Sciences, Beijing, 100080, China}
\author{Franco Nori}
\affiliation{Frontier Research System,  The Institute of
Physical and Chemical Research (RIKEN), Wako-shi, Saitama
351-0198, Japan} \affiliation{CREST, Japan Science and
Technology Agency (JST), Kawaguchi, Saitama 332-0012, Japan}
\affiliation{Center for Theoretical Physics, Physics Department,
Center for the Study of Complex Systems, The University of
Michigan, Ann Arbor, Michigan 48109-1040, USA}
\date{\today}

\begin{abstract}

We study a coupling/decoupling method between a superconducting
qubit and a data bus that uses a controllable time-dependent
electromagnetic field (TDEF). As in recent experiments, the data
bus can be either an LC circuit or a cavity field. When the
qubit and the data bus are initially fabricated, their detuning
should be made far larger than their coupling constant, so these
can be treated as two independent subsystems. However, if a TDEF
is applied to the qubit, then a ``dressed qubit" (i.e., qubit
plus the electromagnetic field) can be formed. By choosing
appropriate parameters for the TDEF, the dressed qubit can be
coupled to the data bus and, thus, the qubit and the data bus
can exchange information with the assistance of the TDEF. This
mechanism allows the scalability of the circuit to many qubits.
With the help of the TDEF, any two qubits can be selectively
coupled to (and decoupled from) a common data bus. Therefore,
quantum information can be transferred from one qubit to
another.

\end{abstract}

\maketitle \pagenumbering{arabic}

\section{Introduction}

Superconducting qubits~\cite{phy} are promising candidates for
quantum information processing and their macroscopic quantum
coherence has been experimentally demonstrated. Single
superconducting qubit experiments also motivate both theorists
and experimentalists to explore the possibility for scaling up
to many qubits.

Two-qubit experiments have been performed in superconducting
charge~\cite{pashkin}, flux~\cite{izmalkov,majer,plourde1}, and
phase qubit~\cite{berkley,xu,mcdermott} circuits. One of the
basic requirements for scalability to many qubits is to
selectively couple any pair of qubits. However, these
experimental
circuits~\cite{pashkin,izmalkov,majer,plourde1,berkley,xu,mcdermott}
are difficult to scale up to many qubits, due to the existence
of the always-on interaction. Theoretical proposals (e.g.,
Refs.~\cite{makhlin,you2,jqprl,you1,
blais,liuyx,zagoskin,migliore, blais1,cleland}) have been put
forward to selectively couple any pair of qubits through a
common data bus (DB).  Some proposals (e.g.,
Refs.~\cite{makhlin,you2,jqprl}) only involve virtual
excitations of the DB modes, while in others (e.g.,
Refs.~\cite{you1, blais,liuyx,zagoskin,migliore,
blais1,cleland}), the DB modes need to be excited. In these
proposals (e.g., Refs. ~\cite{makhlin,you2,jqprl}), the
controllable coupling is implemented by the fast change of the
external magnetic flux, which is a challenge for current
experiments. The switchable coupling between any pair of qubits
can also be implemented by adding additional subcircuits (e.g.,
in Refs.~\cite{averin,plourde}). These additional elements
increase the complexity of the circuits and also might add
additional uncontrollable noise.

Recently, two theoretical approaches~\cite{rigetti,liu1} using
time-dependent electromagnetic fields have been proposed to
control the coupling between two qubits. Both proposals require
that: (i) the detuning between the two qubits is far larger than
their coupling constant; and thus the ratio between the coupling
constant and the detuning is negligibly small. In this case, the
two qubits can be considered as two independent
subsystems~\cite{mcdermott}. (ii) To couple two qubits, the
appropriate time-dependent electromagnetic fields (TDEFs) or
variable-frequency electromagnetic fields must be applied to the
qubits, to achieve coupling/decoupling.

However, there are significant differences between these two
approaches~\cite{rigetti,liu1}. Some are described below.

(i) In the proposal~\cite{rigetti}, two dressed qubits are
formed by the two decoupled qubits and their corresponding
TDEFs. If the parameters of the applied TDEFs are appropriately
chosen so that the transition frequencies of the two dressed
qubits are the same, then the resonant coupling of the two
dressed qubits is realized, and the information between two
decoupled qubits can be exchanged with the help of the TDEFs.
However, for another proposal~\cite{liu1}, one TDEF is enough to
achieve the goal of exchanging information between two decoupled
qubits. This works because there is a nonlinear
coupling~\cite{liu1} between the applied TDEF and the two
decoupled qubits. If the frequency of the applied TDEF is equal
to either the detuning (i.e., the difference) or the sum of the
frequencies of the two qubits, then these two qubits are coupled
to each other and information between these two qubits can be
exchanged.

(ii) For the case in Ref.~\cite{rigetti}, when two qubits are
coupled to each other, the original basis states of each qubit
are mixed by the TDEF, but the frequencies of the qubits remain
unchanged. However, for the proposal in Ref.~\cite{liu1}, both
basis states and transition frequencies of the two qubits remain
unchanged during the coupling/decoupling processes.

The approach in Ref.~\cite{liu1} can be used to scale up to many
qubits by virtue of a common DB~\cite{liu2}, in analogy with
quantum computing with trapped ions~\cite{cirac,sasura,wei1},
and in contrast with the circuit QED approach~\cite{you1,
blais,liuyx,zagoskin,migliore,wallraff,circuitQED}. The
essential differences between the ``trapped ion" proposal for
superconducting qubits~\cite{liu2} and the circuit
QED~\cite{you1,
blais,liuyx,zagoskin,migliore,wallraff,circuitQED} approach are
the following.

(a) When a TDEF is applied to the selected qubit, there are
nonlinear coupling terms~\cite{liu2} between that qubit, the DB
and the TDEF, but these terms do not appear in the circuit QED
proposal~\cite{you1,
blais,liuyx,zagoskin,migliore,wallraff,circuitQED}. This
significant difference provides different coupling mechanisms
for these two proposals.

(b) In Ref.~\cite{liu2}, the frequencies of the qubit and the
data bus always remain unchanged during operations, including
coupling and decoupling. But these frequencies are changed in
the coupling and decoupling stages for the circuit QED (e.g., in
Refs.~\cite{wallraff,circuitQED}).

(c) The qubit-DB coupling is realized in the
proposal~\cite{liu2} by applying a TDEF so that the frequency of
the applied TDEF is equal to the detuning or sum of the
frequencies of the qubit and the data bus; but this qubit-DB
coupling is realized by changing the qubit frequency, so that it
becomes equal to the DB frequency in the circuit QED.

(d) Without an applied TDEF used for the ``trapped ion"
proposal~\cite{liu2}, the qubit and the DB are
decoupled~\cite{liu2}. In contrast, in the circuit QED (e.g., in
Refs.~\cite{wallraff,circuitQED}), the decoupling is realized by
changing the qubit frequency such that the qubit and the DB have
a very large detuning.

From (b), (c), and (d), it is clear that, in circuit QED, the
Hilbert spaces of the qubit and the data bus are always changed
during the coupling/decoupling stage. But, in the ``trapped ion"
approach~\cite{liu2}, the Hilbert spaces of the qubit and the
data bus remain unchanged during the coupling/decoupling
processes .

Also, after our papers in Ref.~\cite{liu1} and Ref.~\cite{liu2}
were submitted, other groups~\cite{bertet,nec} followed our
proposal of using the nonlinear coupling between the qubits and
the TDEF to control the couplings among qubits. Our
approach~\cite{liu1,liu2} works when the frequency of the TDEF
is equal to either the detuning or the sum of the frequencies of
the two qubits (or the qubit and the data bus), then the
coupling between the two qubits is realized, otherwise, these
two qubits are decoupled~\cite{liu1,liu2}.

Motivated by the ``dressed qubits" proposal~\cite{rigetti},  in
this paper, we study how to scale up to many qubits using a
common DB and TDEFs. Our paper is organized as follows. In
Sec.~II, we describe the Hamiltonian of a superconducting flux
qubit coupled to an LC circuit--DB. We explain the decoupling
mechanism using dressed qubits,  and then further explain how
the qubit can be coupled to the DB with the help of the TDEF. In
Sec.~III, the dynamical evolution of the qubit and the data bus
is analyzed. In Sec.~IV, the scalability of our proposed circuit
is discussed. We analyze the implementation of single-qubit and
two-qubit gates with the assistance of the TDEFs. In Sec.~V, we
discuss how to generate entangled states. In Sec.~VI, we use
experimentally accessible parameters to discuss the feasibility
of our proposal.

\section{Dressed states and coupling mechanism}

\subsection{Model}

For simplicity,  we first consider a DB interacting with a singe
qubit. Generally, the DB can represent either a single-mode
light field~\cite{you1, blais,liuyx,zagoskin,migliore}, an LC
oscillator (e.g., Refs.~\cite{liu2,plastina,ntt}),  a large
junction~\cite{buisson,wei,wang,zhou}, or other similar elements
which can be modeled by harmonic oscillators. The qubit can be
either an atom, a quantum dot, or a superconducting quantum
circuit with a Josephson junction---working either in the
charge, phase, or flux regime.

\begin{figure}
\includegraphics[bb=85 320 456 600, width=7 cm, clip]{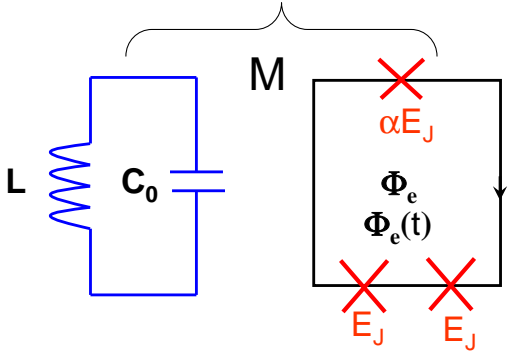}
\caption{(Color online)  A three-junction flux qubit is coupled
to an LC circuit by their mutual inductance $M$. The dc bias
magnetic flux through the qubit is $\Phi_{e}$. A time-dependent
magnetic field $\Phi_{e}(t)$ can be applied to the qubit, so
the qubit can be coupled to the LC circuit. Further details are
explained in the text.} \label{fig1}
\end{figure}

Without loss of generality, we now study a quantum circuit,
shown in Fig.~\ref{fig1}, constructed by a superconducting flux
qubit and an LC circuit acting as a data bus. The interaction
between a single superconducting flux qubit and an LC circuit
has been experimentally realized~\cite{ntt}. The flux qubit
consists of three junctions with one junction smaller by a
factor $0.5 < \alpha <1$ than the other two, identical,
junctions. The LC circuit interacts with the qubit through their
mutual inductance $M$. Then, the total Hamiltonian of the qubit
and the data bus can be written~\cite{liu2,ntt} as
\begin{equation}\label{eq:1}
H_{0}=\hbar\omega a^{\dagger} a +\frac{\hbar}{2} \omega_{q}\,
\sigma_{z}+\hbar\left(\chi\sigma_{+}a+{\rm h.c.}\right),
\end{equation}
in the rotating wave approximation. Here, the qubit operators
are defined by $\sigma_{z}=|e\rangle\langle e|-|g\rangle\langle
g|$, $\sigma_{+}=|e\rangle\langle g|$, and
$\sigma_{-}=|g\rangle\langle e|$;  using its ground $|g\rangle$
and first excited $|e\rangle$ states. The qubit frequency
$\omega_{q}$ in Eq.~(\ref{eq:1}) can be
expressed~\cite{orlando,cyclic} as
\begin{equation*}
\hbar\omega_{q}=2\sqrt{I^2\left(\Phi_{\rm
e}-\frac{\Phi_{0}}{2}\right)^2+T_{RL}^2}
\end{equation*}
with  the bias flux $\Phi_{\rm e}$ and the qubit~\cite{alec}
loop-current $I$. The parameter $T_{RL}$ denotes the tunnel
coupling between the two potential wells of the
qubit~\cite{orlando}. The ladder operators $a$ and $a^{\dagger}$
of the LC circuit are defined by
\begin{subequations}\label{eq:2b}
\begin{eqnarray}
a&=&\sqrt{\frac{C\omega}{\hbar}}\,\varphi
+i \,\sqrt{\frac{1}{\hbar C\omega}} \,Q,\\
a^{\dagger}&=&\sqrt{\frac{C\omega}{\hbar}}\,\varphi -i\,
\sqrt{\frac{1}{\hbar C\omega}}\, Q,
\end{eqnarray}
\end{subequations}
for the magnetic flux $\varphi$ through the LC circuit and the
charge $Q$ stored on the capacitor $C$ of the LC circuit with
the self-inductance $L$. The frequency of the LC circuit is
$\omega=1/\sqrt{LC}$. The magnetic flux $\varphi$ and the charge
$Q$ satisfy the commutation relation $[Q,\, \varphi]=i\hbar$.
The coupling constant $\chi$ between the qubit and the LC
circuit can be written as
\begin{equation*}
\chi=M\sqrt{\frac{\hbar\omega}{2L}} \,\langle e|I|g\rangle.
\end{equation*}

\subsection{Decoupling mechanism between qubit and LC circuit}

Below, we assume that the {\it detuning} $\omega_{q}-\omega$
between the LC circuit and the qubit is larger than their
coupling constant $\chi$, i.e., without loss of generality,
$\omega_{q}-\omega\gg |\chi|$. In this large-detuning condition,
instead of the Hamiltonian $H_{0}$ in Eq.~(\ref{eq:1}), the
dynamical evolution (of the qubit and the LC circuit) is
governed by the effective Hamiltonian~\cite{sun}
\begin{equation}\label{eq:2a}
 H_{0}^{E}=\hbar\omega_{-}\,a^{\dagger}a
+\frac{\hbar}{2}\omega_{q}\sigma_{z}
+\hbar\frac{|\chi|^2}{\omega_{q}-\omega}(1+2a^{\dagger}a)
|e\rangle\langle e|,
\end{equation}
with $\omega_{+}=\omega-|\chi|^2/(\omega_{q}-\omega)$.
Equation~(\ref{eq:2a}) shows that the interaction between the LC
circuit and the  qubit results in a dispersive shift of the
cavity transition or a Stark/Lamb shift of the qubit frequency
$\omega_{q}$. In this large detuning condition, the qubit states
cannot be flipped by virtue of the interaction with the LC
circuit.

Obviously, if the ratio $|\chi|/(\omega_{q}-\omega)$ tends to
zero, then the third term in Eq.~(\ref{eq:2a}) also tends to
zero and $\omega_{-}\approx\omega$. In this limit, the coupling
between the qubit and the data bus can be neglected, that is,
\begin{equation*}
H_{0}^{E}\,\approx\,\hbar\omega a^{\dagger}a
+\frac{\hbar}{2}\omega_{q}\sigma_{z}.
\end{equation*}
The qubit and the LC circuit can be considered as two
independent subsystems, which can be separately controlled or
manipulated. Below, we assume that the LC circuit and the qubit
satisfy the large-detuning condition, e.g.,
$|\chi|/(\omega_{q}-\omega)\sim 0$, when they are initially
fabricated, so they are approximately decoupled.

\subsection{Dressed states}
We now apply a TDEF to the qubit such that the {\it dressed
qubit} can be formed by the applied TDEF and the qubit. Let us
assume that the qubit, driven now by the TDEF, works at the
optimal point~\cite{cyclic}. In this case, the total Hamiltonian
$H$ of the LC circuit and the qubit driven by a TDEF can be
written~\cite{ntt} as
\begin{eqnarray}\label{eq:2}
H&=&H_{0}+\hbar \left(\lambda e^{-i\omega_{c} t} \sigma_{+}+{\rm
H.c.}\right),
\end{eqnarray}
in the rotating wave approximation.  Here $\omega_{c}$ is the
frequency of the TDEF applied to the qubit. $\lambda$ is the
Rabi frequency of the qubit associated with the TDEF. We note
that now the nonlinear coupling strength~\cite{liu2} between the
qubit, LC circuit and the TDEF is zero since the qubit works at
the optimal point.

\begin{figure}
\includegraphics[bb=34 191 584 575, width=9 cm, clip]{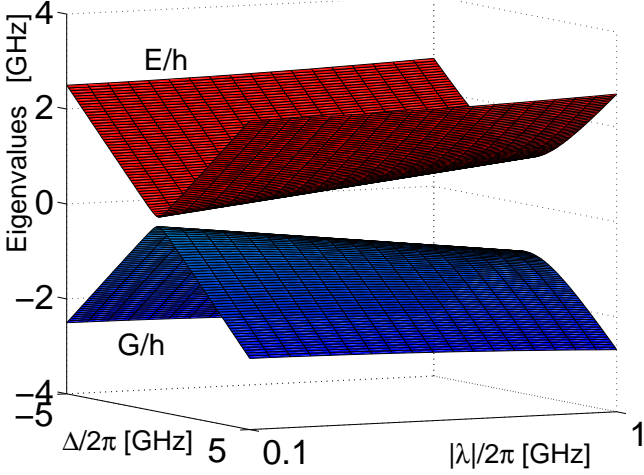}
\caption{(Color online) The dependence of the eigenvalues $E$
and $G$  of Eq.~(\ref{eq:9a}) on the detuning frequency
$\Delta/2\pi=(\omega_{q}-\omega_{c})/2\pi$ and the amplitude of
the Rabi frequency $|\lambda|/2\pi$. Here, the eigenvalues have
been rescaled as frequencies, i.e., $E/h$ and $G/h$. The gap
between these two surfaces corresponds to the frequency
$\Omega/2\pi$ of the dressed qubit.} \label{fig2}
\end{figure}

Since a unitary transformation does not change the eigenvalues
of the system, in the rotating reference frame through a unitary
transformation $U_{R}=\exp(-i\omega_{c}\,\sigma_{z} t/2)$, the
Hamiltonian in Eq.~(\ref{eq:2}) is equivalently transferred to
an effective Hamiltonian
\begin{equation}
H_{e}=U^{\dagger}_{R} H\, U_{R}-i\hbar
U^{\dagger}_{R}\left(\frac{d U_{R}}{d t}\right).
\end{equation}
Hereafter, unless specified otherwise we work in the rotating
reference frame. We can divide the Hamiltonian $H_{e}$ in two
parts, i.e., $H_{e}=H^{(1)}_{e}+H^{(2)}_{e}$ with
\begin{subequations}
\begin{eqnarray}\label{eq:3}
H_{e}^{(1)}&=&\hbar\omega a^{\dagger} a +\hbar \left(\chi
a\,\sigma_{+}\,e^{i\omega_{c}t}+{\rm H.c.}\right), \\
H_{e}^{(2)}&=&\frac{\hbar}{2}\,
\Delta\,\sigma_{z}+\hbar\left(\lambda\,\sigma_{+} +{\rm
H.c.}\right),
\end{eqnarray}
\end{subequations}
where $\Delta=\omega_{q}-\omega_{c}$. The Hamiltonian
$H_{e}^{(2)}$ can be diagonalized and rewritten as
\begin{equation}
H_{e}^{(2)}=\frac{\hbar}{2} \Omega\rho_{z}
\end{equation}
with the transition frequency
\begin{equation*}
\Omega\,=\,\sqrt{(\omega_{q}-\omega_{c})^2+4|\lambda|^2}.
\end{equation*}
Here, $\rho_{z}$ is given by $\rho_{z}=|E\rangle\langle
E|-|G\rangle\langle G|$ in the new basis states
\begin{subequations}\label{eq:6}
\begin{eqnarray}
|E\rangle&=&\cos\frac{\eta}{2}|e\rangle
+e^{i\phi}\sin\frac{\eta}{2}|g\rangle,\\
|G\rangle&=&-\sin\frac{\eta}{2}|e\rangle
+e^{i\phi}\cos\frac{\eta}{2}|g\rangle,
\end{eqnarray}
\end{subequations}
with $\eta=\tan^{-1}(2|\lambda|/\Delta)$. The eigenvalues $E$ and
$G$, corresponding to the eigenstates $|E\rangle$ and $|G\rangle$,
are denoted by
\begin{equation}\label{eq:9a}
E\;=\;-G\;=\;\frac{\hbar}{2}\sqrt{\Delta^2+4|\lambda|^2}.
\end{equation}
The phase $\phi$ is related to the Rabi frequency $\lambda$
($\lambda=|\lambda|e^{-i\phi}$), and the phase $\phi$ can be
controlled by the applied TDEF. In Fig.~{\ref{fig2}}, the
dependence of the eigenvalues $E$ and $G$ on the detuning
$\Delta=\omega_{q}-\omega_{c}$  and the amplitude of the Rabi
frequency $|\lambda|$ has been plotted. The gap between these
two surfaces corresponds to the frequency $\Omega/2\pi$ of the
dressed qubit. It clearly shows that $\Omega$ can be changed by
$\omega_{c}$ when $|\lambda|$ and $\omega_{q}$ are given, and
also changed by $|\lambda|$ when $\omega_{c}$ and $\omega_{q}$
are given. The larger of $|\lambda|$ and $\Delta$ corresponds to
the larger transition frequency $\Omega$ of the dressed qubit.

In fact, the states $|E\rangle$ and $|G\rangle$ could be
interpreted as the {\it dressed states} of the qubit and the
TDEF~\cite{cohen}. Usually, the applied TDEF is considered as in
a coherent state~\cite{cohen}, e.g., $|\alpha \exp(-i\omega_{c}
t)\rangle$. If the creation and annihilation operators of the
TDEF are represented by $b^{\dagger}$ and $b$, then the state
$|\alpha\exp(-i\omega_{c} t)\rangle$ is an eigenstate of the
operator $b$ with the eigenvalue $\alpha \exp(-i\omega_{c} t)$.
The average photon number $\overline{N}$ of the TDEF in this
coherent state is $\overline{N}=|\alpha|^2$, and the width
$\delta \overline{N}$ of the number distribution of photons for
the applied TDEF is $\delta \overline{N}=|\alpha|$.

In the limit $\overline{N}\gg \delta \overline{N} \gg 1$,  the
photon number, absorbed and emitted by the qubit, is negligibly
small, and the qubit is always assumed to be subjected to the
same intensity $|\alpha|^2$ of the applied TDEF during the
operation. Therefore, the TDEF operators $b$ and $b^\dagger$ can
be replaced by the classical number $\alpha\exp(-i\omega_{c} t)$
and its complex conjugate. The relation between $\alpha$ and the
Rabi frequency $\lambda$ of the qubit associated with the
applied TDEF is $|\lambda|\propto |\alpha|$. And also the
coherent state $|\alpha \exp(-i\omega_{c} t)\rangle$,
representing the TDEF, and the qubit state can always be
factorized at any time.

In the rotating reference frame, the coherent state of the TDEF
is always $|\alpha\rangle$, and the dressed qubit-TDEF states
$|E\rangle_{D}$ and $|G\rangle_{D}$ can be understood as the
product state of $|E\rangle$ (or $|G\rangle$) and $|\alpha
\rangle$, that is,
\begin{subequations}
\begin{eqnarray}
|E\rangle_{D}&=&\left(e^{i\phi}\sin\frac{\eta}{2}|g\rangle+\cos\frac{\eta}{2}|e\rangle
\right)|\alpha\rangle=|E\rangle|\alpha\rangle,\\
|G\rangle_{D}&=&\left(e^{i\phi}\cos\frac{\eta}{2}|g\rangle
-\sin\frac{\eta}{2}|e\rangle\right)|\alpha\rangle=|G\rangle|\alpha\rangle.
\end{eqnarray}
\end{subequations}
Therefore, the photon state $|\alpha\rangle$ of the TDEF is
usually omitted when the dressed states are constructed by the
qubit and the TDEF, e.g., in Eq.~(\ref{eq:6}). Hereafter, in
contrast to the dressed states $|G\rangle$ and $|E\rangle$,
$|g\rangle$ and $|e\rangle$ are called ``bare" or undressed
qubit states.

If we assume $\omega >\omega_{c}$, then in the dressed-state
basis of Eq.~(\ref{eq:6}), the effective Hamiltonian $H_{e}$ can
be rewritten as
\begin{equation}\label{eq:9}
H_{e}=\hbar\,\omega a^{\dagger} a+\frac{\hbar}{2}\Omega\rho_{z}+
\hbar (\kappa\, \rho_{-} a^{\dagger}e^{-i\omega_{c}t} +{\rm H.c.})
\end{equation}
with the coupling constant
\begin{equation*}
\kappa=\chi\cos^2(\eta/2)
\end{equation*}
between the dressed qubit and the LC circuit--DB. The ladder
operator $\rho_{-}$ is defined as $\rho_{-} =|G\rangle\langle
E|$. Here, the terms $(\chi/2)\,\sin(\eta)\,\rho_{z}\,a\,
e^{i\omega_{c}t}$, $\sin^2(\eta/2)\,\rho_{-}\, a
\,e^{-i\omega_{c}t}$, and their complex conjugates have been
neglected because of the following reason: there is no way to
conserve energies in these terms, and then they can be neglected
by using the usual rotating-wave approximation.

\subsection{Coupling mechanism between qubit and LC circuit}

To better understand the coupling mechanism,  we can rewrite the
Hamiltonian of Eq.~(\ref{eq:9}), in the interaction picture, as
\begin{equation}\label{eq:10}
H_{e,{\rm int}}=\hbar \kappa\, \rho_{-}
a^{\dagger}e^{i[(\omega-\omega_{c})-\Omega]t} +{\rm H.c.}\,.
\end{equation}
Obviously, the condition
\begin{equation}\label{eq:10a}
\Omega=\omega-\omega_{c}
\end{equation}
is satisfied when the fast oscillating factor
$e^{i[(\omega-\omega_{c})-\Omega]t}$ and its complex conjugate
are always one. In this case, the Hamiltonian (\ref{eq:10})
becomes
\begin{equation}\label{eq:10b}
H_{e,{\rm int}}=\hbar \kappa\, \rho_{-} \,a^{\dagger}+{\rm
H.c.}\,.
\end{equation}

The resonant condition in Eq.~(\ref{eq:10a}) can always be
satisfied by choosing the appropriate frequency $\omega_{c}$ of
the TDEF and the Rabi frequency $\lambda$. Therefore, the
dressed qubit can resonantly interact with the LC circuit, and
then the information can be exchanged between the qubit and the
LC circuit with the help of the TDEF.

\subsection{An example of coupling between a qubit and an LC circuit}

We now numerically demonstrate  the coupling and decoupling
mechanism. For example, let us consider a qubit with frequency
$\omega_{q}/2\pi=2$ GHz which works at the optimal point; the
frequency of the LC circuit is $\omega/2\pi\sim 4$ GHz; and the
coupling constant $|\chi|$ between the LC circuit and the qubit
is $200$ MHz. In this case, the ratio
$|\chi|/(\omega-\omega_{q})=0.1$, and the Stark/Lamb shift for
the qubit frequency is about $20$ MHz, which is much smaller
than the qubit frequency of $2$ GHz. In this case, the
interaction between the qubit and the LC circuit only results in
an ac Stark/Lamb shift, but cannot make qubit states flip.

\begin{figure}
\includegraphics[bb=35 279 537 638, width=8.4 cm, clip]{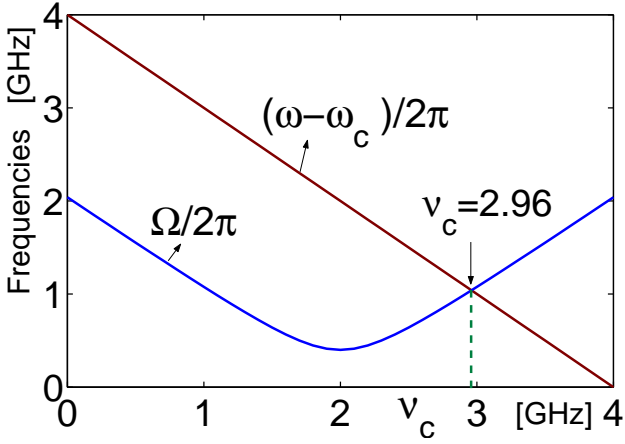}
\caption{(Color online) The frequency $\Omega/2\pi$ of a dressed
qubit (blue curve) and the frequency difference
$(\omega-\omega_{c})/2\pi$ (dark red line); both versus the
frequency $\nu_{c}=\omega_{c}/2\pi$ of the TDEF  for the qubit
frequency $\omega_{q}/2\pi=2$ GHz. The frequency of the LC
circuit is $\omega/2\pi=4$ GHz; and the Rabi frequency of the
qubit associated with the TDEF is $|\lambda|/2\pi=0.2$ GHz. The
crossing point denotes the value of $\nu_{c}\approx 2.96$ GHz
when the condition $\Omega=\omega-\omega_{c}$ is satisfied.}
\label{fig3}
\end{figure}

If we apply a TDEF such that the frequency $\Omega$ of the
dressed qubit satisfies the condition in Eq.~(\ref{eq:10a}),
then the qubit states can be flipped by the interaction with the
LC circuit with the help of the TDEF. In Fig.~\ref{fig3}, the
transition frequency of the dressed qubit
\begin{equation*}
\Omega=\sqrt{(\omega_{q}-\omega_{c})^2+4|\lambda|^2}
\end{equation*}
and the frequency difference $(\omega-\omega_{c})/2\pi$  are
plotted as a function of the frequency $\nu_{c}=\omega_{c}/2\pi$
of the TDEF for the above given frequencies of the qubit and the
LC circuit when the Rabi frequency of the qubit associated with
the TDEF $|\lambda|/2\pi=0.2$ GHz. Figure~\ref{fig3} shows that
when the frequency of the TDEF is
$\nu_{c}=\omega_{c}/2\pi\approx 2.96$ GHz, the condition
$\Omega=\omega-\omega_{c}$ satisfied, and then the qubit is
coupled to the LC circuit with the assistance of the TDEF.
Threfore, the qubit state can be flipped by virtue of the
interaction with the LC circuit with the help of the TDEF.

This could be compared with the switchable coupling circuits in
Ref.~\cite{liu2}, where the qubit basis is always kept in
$\{|g\rangle, \,|e\rangle \}$ no matter if the qubit is coupled
to or decoupled from the LC circuit.  Here, the qubit basis
states (e.g., $|g\rangle$ and $|e\rangle$) will be mixed as in
Eq.~(\ref{eq:6}) in the process of the TDEF-assisted qubit and
LC circuit coupling.

\section{Dynamical evolution of a dressed qubit interacting with
an LC circuit}

\subsection{Resonant Case}

According to the above discussions, the information of the qubit
can be transferred to the data bus with the assistance of an
appropriate TDEF. For convenience, we observe the total system
in another rotating reference frame
$V_{R}=\exp(i\omega_{c}\,\rho_{z}\,t/2)$, then an effective
Hamiltonian from Eq.~(\ref{eq:9}) can be obtained as
\begin{equation}\label{eq:13a}
H^{R}_{e}=\hbar\,\omega a^{\dagger}
a+\frac{\hbar}{2}(\Omega+\omega_{c}) \rho_{z}+ \hbar (\kappa\,
\rho_{-} a^{\dagger} +{\rm H.c.})\,.
\end{equation}
Therefore, the condition of resonant interaction between the
dressed qubit and the data bus is $\Omega=\omega-\omega_{c}$, as
obtained in Eq.~(\ref{eq:10a}). Here we need to emphasize that
the basis states have been changed to $\{|G\rangle, \,|E\rangle
\}$ when the qubit is coupled to the LC circuit with the
assistance of the TDEF, but the qubit basis states are
$\{|g\rangle, \,|e\rangle \}$ when the qubit is decoupled from
the LC cirucit.

According to the Hamiltonian in Eq.~(\ref{eq:13a}), if the LC
circuit and the qubit are initially in the state $|0,
G\rangle=|0\rangle\otimes |G\rangle$, or
$|n+1,\,G\rangle=|n+1\rangle\otimes|G\rangle$, or
$|n,\,E\rangle=|n\rangle\otimes|E\rangle$, then they can evolve
to the following states
\begin{subequations}\label{eq:11}
\begin{eqnarray}
|0,\,G\rangle &\rightarrow&  |0,\,G\rangle,\\
|n,\,E\rangle&\rightarrow& A(t)\left[\cos(\nu
t)|n,\,E\rangle \right.\nonumber\\
&-&\left.ie^{i\delta}\sin(\nu t)|n+1,\,G\rangle\right],\\
|n+1,\,G\rangle&\rightarrow& A(t)\left[\cos(\nu
t)|n+1,\,G\rangle \right. \nonumber\\
&-&\left. ie^{-i\delta}\sin(\nu t)|n,\,E\rangle\right],
\end{eqnarray}
\end{subequations}
with $A(t)=\exp[-i(2n+1)\omega t/2]$, $\nu=|\kappa|\sqrt{n+1}$,
and $\kappa=|\kappa|e^{i\delta}
=|\chi\cos^2(\eta/2)|e^{i\delta}$. It is obvious that the phase
$\delta$ is determined by the coupling constant $\chi$ between
the qubit and the LC circuit. Here, we note that the state
$|n,\,E\rangle$ (or $|m,\,G\rangle$) denotes that the LC circuit
is in the number state $|n\rangle$ (or $|m\rangle$), but the
qubit is in the dressed state $|E\rangle$ (or $|G\rangle$).

In the following discussions, we focus on the case where the LC
circuit is initially in a state $|0\rangle$ or $|1\rangle$.
According to Eq.~(\ref{eq:11}), we can obtain the following
transformations
\begin{subequations}\label{eq:12}
\begin{eqnarray}
|0,\,G\rangle &\rightarrow&  |0,\,G\rangle,\\
|0,\,E\rangle&\rightarrow& B(t)\left[\cos(|\kappa|
t)|0,\,E\rangle \right.\nonumber\\
&-&\left.ie^{i\delta}\sin(|\kappa| t)|1,\,G\rangle\right],\\
|1,\,G\rangle&\rightarrow& B(t)\left[\cos(|\kappa|
t)|1,\,G\rangle \right.  \nonumber\\
&-&\left. ie^{-i\delta}\sin(|\kappa| t)|0,\,E\rangle\right],
\end{eqnarray}
\end{subequations}
with $B(t)=\exp[-i\omega t/2]$.

\subsection{Nonresonant case}
Above, we assumed that the detuning $|\omega_{q}-\omega|$
between the qubit and the LC circuit is far larger than their
coupling $|\chi|$, i.e., $|\chi|/(\omega_{q}-\omega)\sim 0$, and
thus the qubit and the LC circuit are independent. Here we
consider another nonresonant case between the dressed qubit and
the LC circuit in Eq.~(\ref{eq:13a}). We assume that the
detuning $\Lambda$ between the dressed qubit and the LC circuit
satisfies the condition $|\kappa|/\Lambda \ll 1$ with
$\Lambda=\Omega+\omega_{c}-\omega$, but the ratio
$|\kappa|/\Lambda$ does not tend to zero. In this case, the
dynamical evolution of the dressed qubit and the LC circuit is
governed by an effective Hamiltonian
\begin{equation}\label{eq:14a}
 H^{D}_{e}=\hbar\omega_{-}a^{\dagger}a+\frac{\hbar}{2}\Omega^{\prime}\rho_{z}
 +\hbar\frac{|\kappa|^2}{\Lambda}(1+2a^{\dagger}a)
 |E\rangle\langle E|.
\end{equation}
with $\omega_{-}=\omega-(|\kappa|^2/\Lambda)$ and
$\Omega^{\prime}=\Omega+\omega_{c}$. Because the ratio
$|\kappa|/\Lambda$ is not negligibly small, the Stark/Lamb shift
of the dressed qubit frequency or the dispersive shift of the
frequency of the LC circuit should be taken into account.

If the LC circuit and the dressed qubit are initially in states
$|0, G\rangle$, $|1,\,G\rangle$, $|0,\,E\rangle$, or
$|1,\,E\rangle$, then they evolve as follows:
\begin{eqnarray}
|0, G\rangle&\Rightarrow& \exp\left(i\frac{\Omega^{\prime}}{2} t\right)|0, G\rangle, \label{eq:14b}\\
|1,\,G\rangle&\Rightarrow&\exp\left\{i\left(-\omega
+\frac{|\kappa|^2}{\Lambda}
+\frac{\Omega^{\prime}}{2}\right)t\right\}|1,\,G\rangle,\label{eq:14c}\\
|0,\,E\rangle&\Rightarrow&
\exp\left\{-i\left(\frac{\Omega^{\prime}}{2}+\frac{|\kappa|^2}{\Lambda}\right)t\right\}
|0,\,E\rangle,\label{eq:14d}\\
|1,\,E\rangle&\Rightarrow&
\exp\left\{-i\left(\omega+\frac{2|\kappa|^2}{\Lambda}+\frac{\Omega^{\prime}}{2}\right)t\right\}
|1,\,E\rangle \label{eq:14e}.
\end{eqnarray}

\section{Scalable circuit and Quantum Operations}
\subsection{Scalable circuit}
\begin{figure}
\includegraphics[bb=81 361 525 654, width=8 cm, clip]{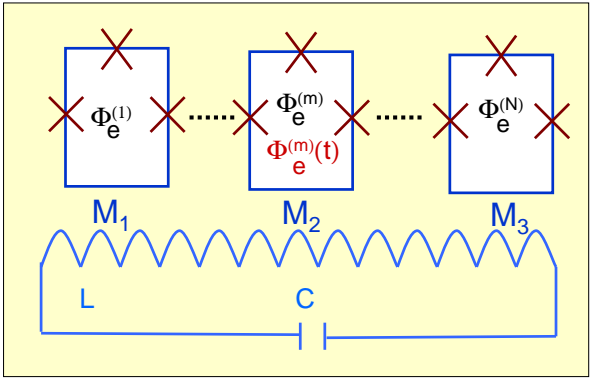}
\caption{(Color online)  The $N$ flux qubits are coupled to an
LC circuit by the their mutual inductances $M_{m}$
($m=1,\,\cdots,\,N$). The bias magnetic flux through the $m$th
qubit is $\Phi_{e}^{(m)}$. A time-dependent electromagnetic
field (TDEF)  can be applied to any one of the qubits (e.g.,
$\Phi_{e}^{(m)}(t)$ through the $m$th qubit) such that the qubit
can be coupled to the LC circuit with the help of the TDEF.}
\label{fig4}
\end{figure}

In the above, we show the basic mechanism of the coupling and
decoupling between a superconducting flux qubit and the LC
circuit. Therefore, a scalable quantum circuit, which is
required for quantum information processing, can be constructed
by $N$ flux qubits and an LC circuit acting as a data bus, shown
in Fig.~\ref{fig4}. The LC circuit interacts with $N$ qubits
through their mutual inductances
$M_{m}\,\,(m=1,\,2,\,\cdots,\,N)$. The distance between any two
nearest qubits is assumed so large that their interaction
through the mutual inductance can be negligibly small. Then, the
total Hamiltonian of qubits and the data bus can be
written~\cite{liu2} as
\begin{equation}\label{eq:1a}
H=\hbar\omega a^{\dagger} a +\frac{\hbar}{2}\sum_{m=1}^{N}
\omega_{m} \sigma^{(m)}_{z}+\hbar\sum_{m=1}^{N}
\left(\chi_{m}\sigma^{(m)}_{+}a+{\rm H.c.}\right)\,,
\end{equation}
in the rotating wave approximation. Here, the $m$th qubit
operators are defined as $\sigma^{(m)}_{z}=|e_{m}\rangle\langle
e_{m}|-|g_{m}\rangle\langle g_{m}|$,
$\sigma^{(m)}_{+}=|e_{m}\rangle\langle g_{m}|$, and
$\sigma^{(m)}_{-}=|g_{m}\rangle\langle e_{m}|$
$(m=1,\,2,\,\cdots,\,N)$ using its ground $|g_{m}\rangle$ and
the first excited $|e_{m}\rangle$ states. The $m$th qubit
frequency $\omega_{m}$ in Eq.~(\ref{eq:1a}) can be
expressed~\cite{orlando} as
\begin{equation*}
\hbar\omega_{m}=2\sqrt{\left[I^{(m)}\left(\Phi^{(m)}_{\rm
e}-\frac{\Phi_{0}}{2}\right)\right]^2+\left[T_{RL}^{(m)}\right]^2}
\end{equation*}
with  the bias flux $\Phi^{(m)}_{\rm e}$ and its loop-current
$I^{(m)}$ of the $m$th qubit~\cite{alec,liu2}. The parameter
$T_{RL}^{(m)}$ denotes the tunnel coupling between two wells in
the $m$th qubit~\cite{orlando}. The ladder operators $a$ and
$a^{\dagger}$ of the LC circuit are defined as in
Eq.~(\ref{eq:2b}). The coupling constant $\chi_{m}$ between the
$m$th qubit and the LC circuit is
\begin{equation*}
\chi_{m}=M_{m}\,\sqrt{\frac{\hbar\omega}{2L}}\, \langle
e_{m}|I^{(m)}|g_{m}\rangle.
\end{equation*}

As in the above discussions, we assume that the detuning
$\omega_{m}-\omega$ between the LC circuit and the $m$th
($m=1,\cdots,N$) qubit is far larger than their coupling
constant $\chi_{m}$. That is, $\chi_{m}/(\omega_{m}-\omega)\sim
0$. Then, all $N$ qubits are decoupled from the LC circuit and
each qubit can be independently manipulated by the TDEF. To
couple a qubit to the LC circuit, an appropriate TDEF is needed
to be applied such that the dressed qubit states can be formed,
and then the dressed qubit can resonantly interact with the LC
circuit.

For convenience, the parameters of any qubit are defined as
follows. The frequency of the TDEF applied to the $m$th qubit is
denoted by $\omega_{m,c}$. The detuning between the $m$th qubit
frequency $\omega_{m}$ and $\omega_{m,c}$ is,
$\Delta_{m}=\omega_{m}-\omega_{m,c}$; $\lambda_{m}$ is the Rabi
frequency of the $m$th qubit associated with the TDEF.
$|G_{m}\rangle$ and  $|E_{m}\rangle$ are eigenstates of the
$m$th qubit with the eigenvalues $G_{m}$ and $E_{m}$.  The
frequency of the $m$th dressed qubit is given by $\Omega_{m}$.
The coupling constant between the $m$th qubit and the LC circuit
is
\begin{equation*}
\kappa_{m}=\chi_{m}\cos^2(\eta_{m}/2),
\end{equation*}
with $\eta_{m}={\rm tan}^{-1} (2|\lambda_{m}|/\Delta_{m})$.

We will now study how to implement the single- and two-qubit
operations for the scalable circuit schematically shown in
Fig.~\ref{fig4}.

\subsection{Single-qubit Operations}

The single-qubit operations of any qubit are easy to implement
by applying a TDEF, which resonantly interacts with the selected
qubit. For instance, if the frequency of the $m$th qubit is
equal to the frequency $\omega_{m,c}$ of the applied TDEF, i.e.,
$ \omega_{m,c}=\omega_{m}$, then the $m$th qubit rotation driven
by the TDEF can be implemented by the single-qubit Hamiltonian
$H_{m,s}$
\begin{equation}\label{eq:13}
H_{m,s}=\hbar|\lambda_{m}|\left(e^{-i\beta_{m}}\sigma^{(m)}_{+}
+e^{i\beta_{m}}\sigma^{(m)}_{-}\right),
\end{equation}
in the rotating reference frame through a unitary transformation
$\exp(-i\omega_{m}\sigma_{z}t)$. Here, the phase $\beta_{m}$ is
determined by the applied TDEF. The time evolution operator of
the Hamiltonian in Eq.~(\ref{eq:13}) can be wrrien as
\begin{equation}\label{eq:14}
U(\theta_{m},\beta_{m})=\exp\left[-i\frac{\theta_{m}}{2}\left(e^{-i\beta_{m}}\sigma^{(m)}_{+}
+e^{i\beta_{m}}\sigma^{(m)}_{-}\right)\right],
\end{equation}
with a duration $t$ and $\theta_{m}=2|\lambda_{m}|t$. Here,
$U(\theta_{m},\beta_{m})$ is a general expression for a
single-qubit operation. This unitary operator
$U(\theta_{m},\beta_{m})$ can be rewritten as a matrix form
\begin{equation}\label{eq:15}
U(\theta_{m},\beta_{m})=\left[\begin{array}{cc}
\cos(\theta_{m}/2)& -ie^{-i\beta_{m}}\sin(\theta_{m}/2)\\
-ie^{i\beta_{m}}\sin(\theta_{m}/2) &\cos(\theta_{m}/2)
\end{array}\right].
\end{equation}
Any single-qubit operation can be derived from
Eq.~(\ref{eq:15}). For instance, a rotation around the $x$ ($y$)
axis can be implemented through Eq.~(\ref{eq:15}) by setting the
applied TDEF such that $\beta_{m}=0$  ($\beta_{m}=\pi/2$). It is
worth pointing out that the operation in Eq.~(\ref{eq:15}) is
defined in the qubit space spanned by $\{|g_{m}\rangle,
\,|e_{m}\rangle \}$.

\subsection{Two-qubit Operations}

To implement two-qubit operations, two qubits should be
sequentially coupled to the LC circuit with the help of the
TDEFs. We now consider how to implement a two-qubit operation
acting on the $m$th and $n$th qubits. For simplicity, we assume
that the classical fields, addressing two qubits to form dressed
states, have the same frequency. Therefore, the following
discussions are confined to the same rotating reference frame.

Let us assume that TDEFs are sequentially applied to the $m$th,
$n$th, and $m$th qubits. The durations of the three pulses are
$\tau_{1}$, $\tau_{2}$, and $\tau_{3}$. After the dressed $m$th
qubit is formed, it can resonantly interact with the LC circuit,
and their dynamical evolution is governed by the Hamiltonian in
Eq.~(\ref{eq:13a}). For the given initial states, they can
evolve as in Eqs.~(\ref{eq:11}) and (\ref{eq:12}). However, for
the dressed $n$th qubit, it does not resonantly interact with
the LC circuit; that is, there is a detuning
$\Lambda_{n}=\Omega_{n}+\omega_{n,c}-\omega$ between the dressed
$n$th qubit and the LC circuit. Their dynamical evolution is
governed by a similar Hamiltonian as in Eq.~(\ref{eq:14a}), with
just replacing the subscript $m$ by $n$, and their states can
evolve as in Eqs.~(\ref{eq:14b}--\ref{eq:14e}).

After these three pulses, the state evolution of two qubits and
the LC circuit can be straightforwardly given~\cite{wei} using
Eqs.~(\ref{eq:12}) and (\ref{eq:14b}) for the total system which
was initially in the state
$|G_{m}\rangle|G_{n}\rangle|0\rangle$, or
$|G_{m}\rangle|E_{n}\rangle|0\rangle$, or
$|E_{m}\rangle|G_{n}\rangle|0\rangle$, or
$|E_{m}\rangle|E_{n}\rangle|0\rangle$.  Here,  e.g., the state
$|G_{m}\rangle|G_{n}\rangle|0\rangle$ denotes that the $m$th and
$n$th qubits are in the states $|G_{m}\rangle$ and
$|G_{n}\rangle$, but the LC circuit is in the state $|0\rangle$.

If the durations $\tau_{1}$ and $\tau_{3}$ of the first and
third pulses applied to the $m$th qubit satisfy the conditions
$\sin(|\kappa_{m}|\tau_{1})=0$ and
$\sin(|\kappa_{m}|\tau_{3})=0$, then the above four different
initial states, e.g., $|G_{m}\rangle|G_{n}\rangle|0\rangle$,
have the following dynamical evolutions
\begin{subequations}
\begin{eqnarray}
|G_{m}\rangle|G_{n}\rangle|0\rangle&\rightarrow&
\exp(i\xi_{1})|G_{m}\rangle|G_{n}\rangle|0\rangle,\\
|G_{m}\rangle|E_{n}\rangle|0\rangle&\rightarrow&
\exp(i\xi_{2})|G_{m}\rangle|E_{n}\rangle|0\rangle,\\
|E_{m}\rangle|G_{n}\rangle|0\rangle&\rightarrow&
\exp(i\xi_{3})|E_{m}\rangle|G_{n}\rangle|0\rangle,\\
|E_{m}\rangle|E_{n}\rangle|0\rangle&\rightarrow&
\exp(i\xi_{4})|E_{m}\rangle|E_{n}\rangle|0\rangle
\end{eqnarray}
\end{subequations}
with
\begin{subequations}
\begin{eqnarray}
\xi_{1}&=&\frac{\Omega_{n}}{2}\tau_{2}\\
\xi_{2}&=&-\xi_{1}
-\frac{|\kappa_{n}|^2}{\Lambda_{n}}\tau_{2},\\
\xi_{3}&=&-\omega(\tau_{1}+\tau_{3})+\xi_{1},\\
\xi_{4}&=&-\omega(\tau_{1}+\tau_{3})+\xi_{2}.
\end{eqnarray}
\end{subequations}
 Here, we neglect the free
evolution of another uncoupled qubit when one qubit is coupled
to the LC circuit. After the above three pulses with the given
durations, a two-qubit phase operation $U_{mn}$ can be
implemented in the basis of the two-qubit dressed states
$\{|E_{m}\rangle|E_{n}\rangle,\, |E_{m}\rangle|G_{n}\rangle,
\,\,|G_{m}\rangle|E_{n}\rangle,\,\,
|G_{m}\rangle|G_{n}\rangle\}$. The matrix form of the operation
$U_{mn}$ is
\begin{equation}
U_{mn}=\left(\begin{array}{cccc} e^{i\xi_{4}}&0&0&0\\
0&e^{i\xi_{3}}&0&0\\
0&0&e^{i\xi_{2}}&0\\
0&0&0& e^{i\xi_{1}}\end{array}\right).
\end{equation}
We note that this two qubit operation is in the rotating
reference frame. Of course, it is also easy to obtain a
two-qubit operation in the bare (undressed) basis
$\{|g_{m}\rangle|g_{n}\rangle,\, |g_{m}\rangle|e_{n}\rangle,
\,\,|e_{m}\rangle|g_{n}\rangle,\,\,
|e_{m}\rangle|e_{n}\rangle\}$ by applying the single-qubit
operations on the $m$th and $n$th qubits separately. The
single-qubit operations can be given by choosing the appropriate
parameters in Eq.~(\ref{eq:15}) for a general expression of the
single-qubit operations.

A two-qubit operation and single-qubit rotations are needed for
universal quantum computing~\cite{divincenzo}. Therefore, the
two-qubit operation $U_{mn}$, accompanied by arbitrary
single-qubit rotations, Eq.~(\ref{eq:15}), of the $m$th and
$n$th qubits, forms a universal set.

\section{Generation of entangled states}

In this section, we will study how to generate an entangled
state between any two qubits, e.g., $m$th and $n$th qubits, with
the assistance of TDEFs.

We assume that the qubits are initially prepared in the dressed
states, e.g, $|E_{m}\rangle\otimes|G_{n}\rangle$, but the LC
circuit is initially in its ground state $|0\rangle$. In this
case, we can apply two pulses to generate an entangled state.
The first pulse with the frequency $\omega_{m,c}$ brings the
$m$th qubit to resonantly interact with the LC circuit. The
interaction Hamiltonian is described by Eq.~(\ref{eq:13a}) in
the rotating reference frame. But there is no interaction
between the LC circuit and the $n$th qubit. With the pulse
duration $\tau_{1}$, the state
$|E_{m}\rangle\otimes|G_{n}\rangle\otimes |0\rangle\equiv
|E_{m}, G_{n}, 0\rangle$ evolves to the state
\begin{eqnarray}\label{eq:23}
|\psi(\tau_{1})\rangle&=&\cos(|\kappa_{m}|
\tau_{1})|E_{m},\, G_{n},\,0\rangle\nonumber\\
&-&ie^{i\delta_{m}}\sin(|\kappa_{m}| \tau_{1})|G_{m},\,
G_{n},\,1\rangle,
\end{eqnarray}
which can be written as
\begin{eqnarray}\label{eq:24}
|\psi^{\prime}(\tau_{1})\rangle&=&e^{-i\omega_{m,c}\tau_{1}/2}\cos(|\kappa_{m}|
\tau_{1})|E_{m},\, G_{n},\,0\rangle\\
&-&ie^{i\delta_{m}}e^{i\omega_{m,c}\tau_{1}/2}\sin(|\kappa_{m}|
\tau_{1})|G_{m},\, G_{n},\,1\rangle \nonumber
\end{eqnarray}
after removing the rotating reference frame. Here, the global
phase factor $e^{-i(\omega_{n}+\omega)\tau_{1}/2}$ has been
neglected and $\delta_{m}$ is given by
$\kappa_{m}=|\kappa_{m}|e^{i\delta_{m}}$.

After the first pulse, the second pulse assists the $n$th qubit to
resonantly interact with the LC circuit. With the duration
$\tau_{2}$ of the second pulse, the state $|\psi(\tau_{1})\rangle$
will evolve to the state
\begin{eqnarray}\label{eq:25}
|\psi(\tau_{2})\rangle&=&e^{i\vartheta_{1}}\cos(|\kappa_{m}|
\tau_{1})|E_{m}, G_{n},\,0\rangle \\
&-& ie^{i\delta_{m}}e^{i\vartheta_{2}}\sin(|\kappa_{m}|
\tau_{1})\cos(|\kappa_{n}|
\tau_{2})|G_{m},\,G_{n},\,1\rangle \nonumber\\
&-&
e^{i(\delta_{m}-\delta_{n})}e^{i\vartheta_{3}}\sin(|\kappa_{m}|
\tau_{1})\sin(|\kappa_{n}| \tau_{2})|G_{m},\,E_{n},\,0\rangle.
\nonumber
\end{eqnarray}
after removing the rotating reference frame. Here $\delta_{n}$ is
determined by $\kappa_{n}=|\kappa_{n}|e^{-i\delta_{n}}$, and
\begin{subequations}\label{eq:26}
\begin{eqnarray}
\vartheta_{1}&=&\frac{1}{2}[-\omega_{m,c}\tau_{1}+(\omega_{n,c}-\omega_{m})
\tau_{2}],\\
\vartheta_{2}&=&\frac{1}{2}[\omega_{m,c}\tau_{1}+(\omega_{m}-\omega+\omega_{n,c})
\tau_{2}],\\
\vartheta_{3}&=&\frac{1}{2}[\omega_{m,c}\tau_{1}+(\omega_{m}-\omega-\omega_{n,c})
\tau_{2}].
\end{eqnarray}
\end{subequations}

If the duration $\tau_{2}$ of the second pulse is chosen such that
$\cos(|\kappa_{n}| \tau_{2})=0$, then an entangled state
$|\psi_{E}\rangle$ is created as
\begin{eqnarray}\label{eq:27}
|\psi_{E}\rangle&=&e^{i\vartheta_{1}}\cos(|\kappa_{m}|
\tau_{1})|\,E_{m},
G_{n},0\rangle \nonumber\\
&-&e^{i(\delta_{m}-\delta_{n})}e^{i\vartheta_{3}}\sin(|\kappa_{m}|
\tau_{1})|\,G_{m},E_{n},0\rangle.
\end{eqnarray}
It is very easy to find that we can prepare different entangled
states by choosing the duration $\tau_{1}$ and the phase
difference $\delta_{m}-\delta_{n}$. For example, if the duration
$\tau_{1}$ for the first pulse and the phase difference
$\delta_{m}-\delta_{n}$ are well chosen so that
\begin{equation}
\tau_{1}=\frac{\pi}{4|\kappa_{m}|},
\end{equation}
then we can get maximally entangled states
\begin{equation}\label{eq:37}
|\psi_{M}\rangle=\frac{1}{\sqrt{2}}[|\,E_{m},
G_{n}\rangle-e^{i\delta_{p}} |\,G_{m},E_{n}\rangle].
\end{equation}
with
$\delta_{p}=\delta_{m}-\delta_{n}+\vartheta_{3}-\vartheta_{1}$.
Here, a global phase factor $e^{i\vartheta_{1}}$ has been
neglected. If the condition
\begin{equation}\label{eq:38}
\delta_{p}=\delta_{m}-\delta_{n}+\vartheta_{3}-\vartheta_{1}=2\,l\pi,
\end{equation}
with an integer $l$, can also be further satisfied, then a Bell
state
\begin{equation}\label{eq:39}
|\psi_{B}\rangle=\frac{1}{\sqrt{2}}(|\,E_{m},
G_{n}\rangle-|\,G_{m},E_{n}\rangle)
\end{equation}
can be obtained from Eq.~(\ref{eq:37}). Here, $\delta_{m}$ and
$\delta_{n}$ are determined by the coupling constants between
the qubits and the LC circuit, thus from the experimental point
of view, it cannot be conveniently adjusted. Therefore, once the
amplitudes and frequencies of the two TDEFs are pre-chosen, the
condition in Eq.~(\ref{eq:38}) might not be easy to satisfy.
Thus, the maximally entangled state in Eq.~(\ref{eq:37}) is
easier to generate compared with the Bell state in
Eq.~(\ref{eq:39}). However, it is still possible to adjust the
amplitudes, frequencies and durations of the TDEFs at the same
time to satisfy the condition in Eq.~(\ref{eq:38}), and then the
Bell state in Eq.~(\ref{eq:39}) can be created.

If the two-qubit states are initially in undressed states (e.g.,
$|g_{m}\rangle\otimes|g_{n}\rangle$), then, we need to first
make single-qubit rotations on the two qubits, such that
$|g_{m}\rangle$ and $|g_{n}\rangle$ can be rotated to
$|E_{m}\rangle$ and $|G_{n}\rangle$, respectively. After this
two single-qubit rotations, we repeat the above steps to obtain
Eqs.~(\ref{eq:23}) and (\ref{eq:25}). Then, we can get an
entangled state, which is the same as in Eq.~(\ref{eq:27})
except that the phases are different from $\vartheta_{1}$,
$\vartheta_{3}$, $\delta_{m}$, and $\delta_{n}$. To prepare
entangled states with the undressed qubit states, we need to
make another two single qubit operations such that
$|G_{m}\rangle$ ($|E_{m}\rangle$) and $|G_{n}\rangle$
($|E_{n}\rangle$) change to $|g_{m}\rangle$ ($|e_{m}\rangle$)
and $|g_{n}\rangle$ ($|e_{n}\rangle$). Then entangled bare qubit
states can be obtained.

\section{Discussions on experimental feasibility}
As an example of superconducting flux qubits interacting with an
LC circuit, we show a general method on how to scale up many
qubits using dressed states. We further discuss two
experimentally accessible superconducting circuits with the
given parameters.

\subsection{Flux qubit interacting with an LC circuit}

\begin{figure}
\includegraphics[bb=30 184 555 590, width=8.0cm, clip]{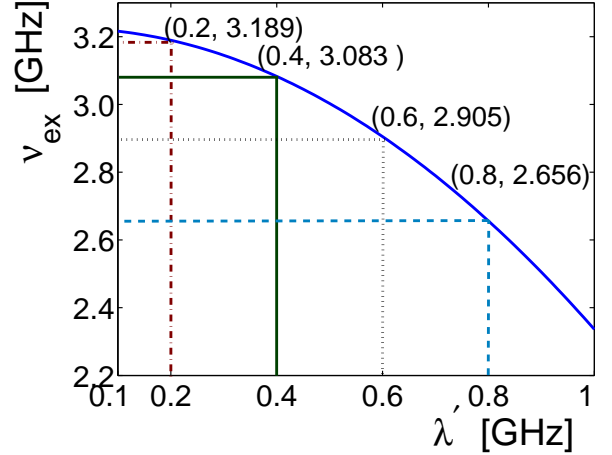}
\caption{(Color online)  To couple qubit to the LC circuit, the
frequency $\nu_{ex}$, applied to the qubit, is plotted as a
function of the coupling constant $\lambda^{\prime}$ between the
qubit and the external microwave. As example, four different
points are marked in the curve to show the required frequencies
of external microwave when different coupling constants
$\lambda^{\prime}$ are given.} \label{fig5}
\end{figure}

A recent experiment~\cite{ntt} has demonstrated the Rabi
oscillations between a single flux qubit and a superconducting
LC circuit. In this experiment~\cite{ntt}, the coupling constant
$g^{\prime}$ between the qubit and the LC circuit is about
$g^{\prime}=0.2$ GHz, the qubit frequency $\nu_{q}$ at the
optimal point is $\nu_{q}=2.1$ GHz, and the frequency of the LC
circuit is $\nu_{LC}=4.35$ GHz.  So the frequency difference
between the LC circuit and the qubit is $\nu_{LC}-\nu_{q}=2.25$
GHz. The ratio $g^{\prime}/(\nu_{LC}-\nu_{q})$ of the coupling
constant $g^{\prime}$ over the frequency difference
$\nu_{LC}-\nu_{q}$ is about $0.089$. Therefore, the dispersive
shift (or Lamb shift) of the LC circuit (qubit) due to
nonresonantly interaction with the qubit (LC circuit) is about
$0.018$ GHz.

If a TDEF with the frequency $\nu_{\rm ex}$ is applied to the
qubit, then the TDEF and the qubit can form a dressed qubit with
the frequency
\begin{equation}
\nu_{D}=\sqrt{(\nu_{\rm
ex}-\nu_{q})^2+4|\lambda^{\prime}|^2}\,.
\end{equation}
Here, the coupling constant between the qubit and the TDEF is
$\lambda^{\prime}$. When the frequency $\nu_{D}$ of the dressed
qubit satisfies the condition
\begin{equation}\label{eq:35}
\nu_{D}=\nu_{LC}-\nu_{\rm ex},
\end{equation}
as shown in Eq.~(\ref{eq:10a}), then the dressed qubit can be
resonantly coupled to the LC circuit. From the condition in
Eq.~(\ref{eq:35}), we derive another equation
\begin{equation}\label{eq:36}
\nu_{\rm
ex}=\frac{1}{2}(\nu_{LC}+\nu_{q})-\frac{2|\lambda^{\prime}|^2}{
(\nu_{LC}-\nu_{q})}.
\end{equation}
To make the dressed qubit couple to the LC circuit,
Eq.~(\ref{eq:36}) shows that we should choose the different
external frequencies $\nu_{\rm ex}$ for different coupling
constants $\lambda^{\prime}$ when the frequencies of the data
bus $\nu_{LC}$ and the qubit $\nu_{q}$ are given.

In Fig.~\ref{fig5}, the frequency $\nu_{\rm ex}$ is plotted as a
function of $\lambda^{\prime}$, which is in the interval
$0.1\sim 1$ GHz, for the above given frequencies of the qubit
and LC circuit. Figure~\ref{fig5} clearly shows that the
frequencies $\nu_{\rm ex}$ of the applied external microwave are
different for the different $\lambda^{\prime}$ in order to
couple the qubit to the LC circuit with the assistance of the
TDEF. As an example, four different points are marked in
Fig.~\ref{fig5} to show the required frequencies $\nu_{\rm ex}$
of external microwave when the coupling constants
$\lambda^{\prime}$ are different. For example, if
$\lambda^{\prime}=0.4$ GHz, then the applied external microwave
should have the frequency $\nu_{\rm ex}=3.083$ GHz to make the
dressed qubit resonantly couple to the LC circuit. And then the
effective coupling constant between the dressed qubit and the LC
circuit is about $\kappa^{\prime}\approx 0.1996$ GHz for the
coupling constant $g^{\prime}=0.2$ GHz between the qubit and the
LC circuit. Therefore the microwave assisted resonant
interaction between the qubit and the LC circuit can be realized
in the current experimental setup~\cite{ntt}. Furthermore, this
circuit can be also scaled up to many qubits.

\subsection{Charge qubit interacting with a single-mode
cavity field}

We consider another experimental example of a charge qubit
interacting with a single-mode cavity field~\cite{wallraff}. In
this experiment~\cite{wallraff}, the qubit frequency $\nu_{q}$
is about $8$ GHz at the degeneracy point. The frequency
$\nu_{c}$ of the cavity field is about $6$ GHz. The coupling
constant $g^{\prime}$ between the qubit and the cavity field can
be, e.g., $50$ MHz, then the ratio between $g^{\prime}$ and the
detuning $\nu_{q}-\nu_{c}$ is
$g^{\prime}/(\nu_{q}-\nu_{c})=0.025$. This means that the qubit
and the cavity field is in the large detuning regime.

If an ac electric field with frequency $\nu_{\rm ex}$ is applied
to the gate of the charge qubit, then the qubit and the ac field
can together form a dressed qubit.  If we choose appropriate
parameters for the ac field, the dressed qubit can be resonantly
coupled to the cavity field and then the qubit and the cavity
field can exchange information with the assistance of the ac
field. For example, if the Rabi frequency of the qubit
associated with the ac field is about, e.g.,
$\lambda^{\prime}\sim 100$ MHz, then the dressed qubit can be
resonantly coupled to the cavity field  when the frequency
$\nu_{ex}$ of the ac field is $7.01$ GHz, which is obtained from
Eq.~(\ref{eq:36}).

For the above two examples, we need to stress that the bias of
the ``bare" charge or flux qubit is always kept to the optimal
point during the operations. In the coupling process, the
external microwave mixes the two ``bare" qubit states, and the
dressed qubit states are resonantly coupled to the data bus (an
LC circuit or a single-mode cavity field).

\section{Conclusion}
In conclusion, using an example of a superconducting flux qubit
interacting with an LC circuit--data bus, we study a method to
couple and decouple selected qubits with the data bus. This
method can be realized with the assistance of time-dependent
electromagnetic fields (TDEFs). If a TDEF is applied to a
selected qubit, then dressed qubit-TDEF states can be formed. By
choosing appropriate parameters of the TDEF, the dressed qubit
can interact resonantly with the data bus. However, when the
TDEF is removed, then the qubit and the data bus are decoupled.
By using this mechanism, many qubits can be selectively coupled
to a data bus. Thus, quantum information can be transferred from
one qubit to another through the data bus with the assistance of
the TDEF.

We stress the following: (i) all qubits are decoupled from the
data bus when their detunings to the data bus are far larger
than their coupling constants to the LC circuit; (ii) the qubits
can be independently manipulated by the TDEFs resonantly
addressing them (for example, if $\omega_{m,c}=\omega_{m}$, then
the $m$th qubit is addressed by its TDEF); (iii) to couple any
one of the qubits to the data bus, an appropriate TDEF is needed
to be applied such that the dressed qubit can be resonantly
coupled to the data bus, and then the information of the qubit
can be transferred to the data bus with the help of the TDEF.

We emphasize that all superconducting qubits (charge or flux
qubits) can work at their optimal points during the coupling and
decoupling processes with the assistance of the TDEF. Although
this coupling/decoupling mechanism is mainly focused on
superconducting flux qubits, it can also be applied to either
charge (e.g, in Refs.~\cite{you1,wallraff}), or
phase~\cite{martinis} qubits, as well as other solid state
systems. For instance, the coupling between two quantum-dot
qubits can be switched on and off by using this method, or a
large number of quantum-dot qubits can be scaled up by using a
single-mode electromagnetic field with the assistance of the
TDEF.

\section{acknowledgments}
This work was supported in part by the National Security Agency
(NSA) and Advanced Research and Development Activity (ARDA)
under Air Force Office of Research (AFOSR) contract number
F49620-02-1-0334; and also supported by the National Science
Foundation grant No.~EIA-0130383; and also supported by Army
Research Office (ARO) and Laboratory of Physical Sciences (LPS).
The work of C.P. Sun is also partially supported by the NSFC and
FRP of China with No. 2001CB309310.

\end{document}